\documentclass[]{aastex631}
\usepackage{amsmath}
\usepackage{graphicx}
\usepackage{threeparttable}
\usepackage{txfonts}
\usepackage{soul}
\usepackage{bm} 
\usepackage{tabularx} 
\usepackage{booktabs}
\usepackage{varwidth}
\usepackage{graphicx}

\shorttitle{Periodic Emission Frequency Modulation in FRB 20240114A}
\shortauthors{Li et al.}

\begin{document}
\title{Periodic Emission Frequency Modulation in a Hyperactive Fast Radio Burst}

\correspondingauthor{Fa-Yin Wang}
\email{fayinwang@nju.edu.cn}

\author[0009-0007-3326-7827]{Rui-Nan Li}
\affiliation{School of Astronomy and Space Science, Nanjing University Nanjing 210023, China}

\author{Hao-Tian Lan}
\affiliation{School of Astronomy and Space Science, Nanjing University Nanjing 210023, China}

\author{Zhen-Yin Zhao}
\affiliation{School of Astronomy and Space Science, Nanjing University Nanjing 210023, China}

\author{Qin Wu}
\affiliation{School of Astronomy and Space Science, Nanjing University Nanjing 210023, China}

\author[0000-0003-4157-7714]{Fa-Yin Wang}
\affiliation{School of Astronomy and Space Science, Nanjing University Nanjing 210023, China}
\affiliation{Key Laboratory of Modern Astronomy and Astrophysics (Nanjing University) Ministry of Education, China}

\author{Zi-Gao Dai}
\affiliation{Department of Astronomy, University of Science and Technology of China 230026, China}

\begin{abstract}
Fast radio bursts (FRBs) are intense, short-duration radio transients of mysterious origin. They have been detected across a wide range of frequencies from 110 MHz to 8 GHz. Their spectral properties, remaining poorly understood, are essential for understanding the intrinsic radiation mechanism and propagation effects. 
Here, we report the discovery of a periodic modulation in the central emission frequency of FRB 20240114A, based on more than one thousand bursts collected by an ultra-wideband receiving system. The burst central frequencies reveals a significant modulation with a period of $\sim 112$ days. The statistical significance of this detected periodicity exceeds $6\sigma$ for both the Lomb-Scargle and phase-folding methods. Within a single period, the central emission frequency exhibits a systematic drift from lower to higher values. We evaluate several physical mechanisms for this unique spectral evolution. The free-free absorption together with cyclotron resonant absorption in a binary system or free precession models could potentially explain such behavior.
The discovery of this periodic frequency modulation unveils a new layer of complexity in the underlying radiation mechanism and propagation effect of FRBs.
\end{abstract}


\section{Introduction} \label{sec:intro}
Fast radio bursts (FRBs) are detected across a broad spectral window, ranging from as low as 110 MHz \citep{Pleunis2021_180916,Marazuela2021} up to at least 8 GHz \citep{Gajjar2018}. 
Observations show that bursts from repeating FRBs typically exhibit narrow-band spectra, with characteristic fractional bandwidths of $\Delta \nu / \nu \sim 0.3$ \citep{Pleunis2021_cat, Chime/FrbCollaboration2026}. To determine whether this narrow-band feature is an intrinsic property or merely a selection effect caused by limited instrumental bandpasses \citep{Aggarwal2021b}, recent monitoring campaigns have deployed ultra-wideband receivers. Observations utilizing the Effelsberg telescope and the Parkes telescope have shown that while the band-limited nature persists, bursts with fractional bandwidths approaching unity occasionally occur \citep{Kumar2021, Kumar2023,Limaye2025}.

Ultra-wide band monitoring of active repeating FRBs provides a unique opportunity to track variations in spectral properties from burst to burst over a long time spans. As the physical origin of FRBs remains debated, proposed radiation mechanisms are broadly categorized into magnetospheric models and far-away relativistic shock models \citep{Zhang2023Review}. The observed spectrum is a critical diagnostic that depends on both the intrinsic radiation mechanism and extrinsic propagation effects. In the framework of coherent curvature radiation within a magnetosphere, the emission frequency is tied to the curvature radius and the Lorentz factor of particle bunches \citep{Kumar2017,Yang2018}. Conversely, in relativistic shock models invoking synchrotron maser emission, the frequency is governed by the bulk Lorentz factor of the shock and the properties of the upstream medium \citep{Metzger2019,Beloborodov2020}. Furthermore, propagation effects within the local environment, such as free-free absorption and plasma lensing, can significantly modulate the observed spectra \citep{cordes2017,Ioka2020,LiRN2026}. 
If the central engine is a rotating magnetar or resides in an interacting binary system, the dynamic circum-burst environment could imprint periodic signatures onto the observed frequency. Indeed, the search for such periodic behaviors is strongly motivated by the confirmation that FRB 200428 originated from a Galactic magnetar \citep{CHIME2020_1935, Bochenek20201935}, along with 
From observations, some active repeaters are probably hosted in binary systems \citep{Wang2022,Anna-Thomas2023,ZhangB2025,Wang2025,LiYe2026}. To date, periodic modulations have been discovered either in burst activity, as seen in FRB 20180916B \citep{CHIME2020180916} and possibly in FRB 20121102A \citep{Rajwade121102}, or in the variations of the Faraday rotation measure (RM), as candidates in FRB 20201124A and FRB 20220529A \citep{XuJW2025, Liang2025}. These observational signatures are widely interpreted as evidence for the orbital motion, spin or precession of the central compact object \citep{Ioka2020, Beniamini2020, Levin2020}. 
However, the modulations on emission frequency have not been found.

FRB 20240114A is a hyperactive repeating source discovered by the Canadian Hydrogen Intensity Mapping Experiment (CHIME) \citep{Shin2026}. Its high burst rate and prolonged active episodes make it an ideal object for long-term monitoring. Extensive observations by the Five-hundred-meter Aperture Spherical radio Telescope (FAST) have yielded a spectacular sample of over 17,000 bursts \citep{ZhangJS2025}. Meanwhile, ultra-wide band observations of this FRB have been performed by the Parkes telescopes \citep{Uttarkar2026}. In a high-cadence Parkes campaign lasting $\sim16$ months, over 5,526 bursts were detected. Our analysis is based on the Parkes observations. 

In this paper, we report the discovery of a periodic modulation in the central emission frequency of FRB 20240114A, utilizing the high-cadence Parkes sample from \citep{Uttarkar2026}. This paper is organized as follows. The description of the observational data used in this work is provided in Section \ref{sec:data}. The results of the periodicity search using the Lomb-Scargle and phase-folding methods are presented in Section \ref{sec:results}. Possible physical interpretations for the discovered periodic frequency modulation are explored in Section \ref{sec:theory}. We conclude with a discussion in Section \ref{sec:dis} and a brief summary in Section \ref{sec:summary}.

\section{Data Sets} \label{sec:data}
FRB 20240114A was monitored using the 64-m Parkes (Murriyang) radio telescope equipped with the UWL (704-4032 MHz) and MARS (7881-8905 MHz) receivers \citep{Uttarkar2026}. Notably, all detections occurred within the UWL band, though the MARS receiver accounted for only a minor percentage of the total observing time. The $\sim$16-month campaign yielded 5526 bursts with a signal-to-noise ratio (S/N) $\ge 7.5$. Burst identification was executed via the \texttt{heimdall} pipeline across a DM range of $100$--$1200$ $\mathrm{pc\ cm}^{-3}$. The full 3328 MHz UWL bandwidth was dynamically sub-banded into widths ranging from 64 to 3328 MHz to perform the burst searching, and manually identify the emission range for each burst. 

For the purpose of this study, we restrict our sample to bursts with S/N $>$ 20 \citep{Uttarkar2026}. Since our periodicity search relies on accurately characterizing the central frequency of each burst, we restrict the burst sample to $\rm{S/N} > 20$ to minimize measurement errors and ensure a robust result. The identified central frequency is highly sensitive to the exact DM used for de-dispersion. It has been shown that for low-S/N bursts, precisely determining the true DM via structure maximization methods is challenging \citep{Pleunis2021_cat, Uttarkar2026}. Consequently, incorporating low-S/N samples into the analysis could introduce non-negligible noise and artificially degrade the statistical significance of the periodicity. Theoretically, this empirical cut is further supported by the scaling relation for the uncertainty of the measured frequency: $\sigma_{\nu} \propto \frac{\Delta \nu}{\rm{S/N}}$ \citep{Cordes2003}, which explicitly dictates that lower S/N bursts inherently suffer from larger measurement errors.

Over 700 bursts from FRB 20240114A have recently been reported using the Effelsberg 100-m telescope (1.3–6.0 GHz) \citep{Limaye2025}. Although their central frequency evolution is broadly consistent with the Parkes measurements, we do not incorporate the Effelsberg data into our core analysis. The significant differences in telescope sensitivities, effective bandpasses, RFI conditions, and the sparse observational coverage (comprising only four widely separated epochs) could introduce substantial systematic errors.

\section{Searching Periodicity in Frequency} \label{sec:results}
In this section, we employ the Lomb-Scargle periodogram and phase-folding techniques to systematically investigate the periodic modulation of the burst central frequencies.
\subsection{Lomb-Scargle Method}
The Lomb-Scargle method estimates the spectral power over a specified period search range and identifies the period with the highest statistical significance. The spectral power is defined as
\begin{equation}\label{equ:1}
\begin{aligned}
P_{\text{LS}}(\omega) = \frac{1}{2} & \left\{ \frac{\left[ \sum_{i} \text{$\nu$}_i \cos \omega(t_i - \tau) \right]^2}{\sum_{i} \cos^2 \omega(t_i - \tau)} \right. \\
& \left. + \frac{\left[ \sum_{i} \text{$\nu$}_i \sin \omega(t_i - \tau) \right]^2}{\sum_{i} \sin^2 \omega(t_i - \tau)} \right\},
\end{aligned}
\end{equation}
where $\omega$ is the angular frequency, $t_i$ is the arrival time of the $i$-th burst, and $\nu_i$ is its corresponding central frequency. The time-delay parameter $\tau$ is defined for each $\omega$ to ensure the periodogram's invariance under time translations
\begin{equation}\label{equ:2}
\begin{aligned}
\tau = \frac{1}{2\omega} \tan^{-1} \left( \frac{\sum\limits_{i} \sin 2\omega t_i}{\sum\limits_{i} \cos 2\omega t_i} \right),
\end{aligned}
\end{equation}

\begin{figure} 
\centering
\includegraphics[width=120 mm]{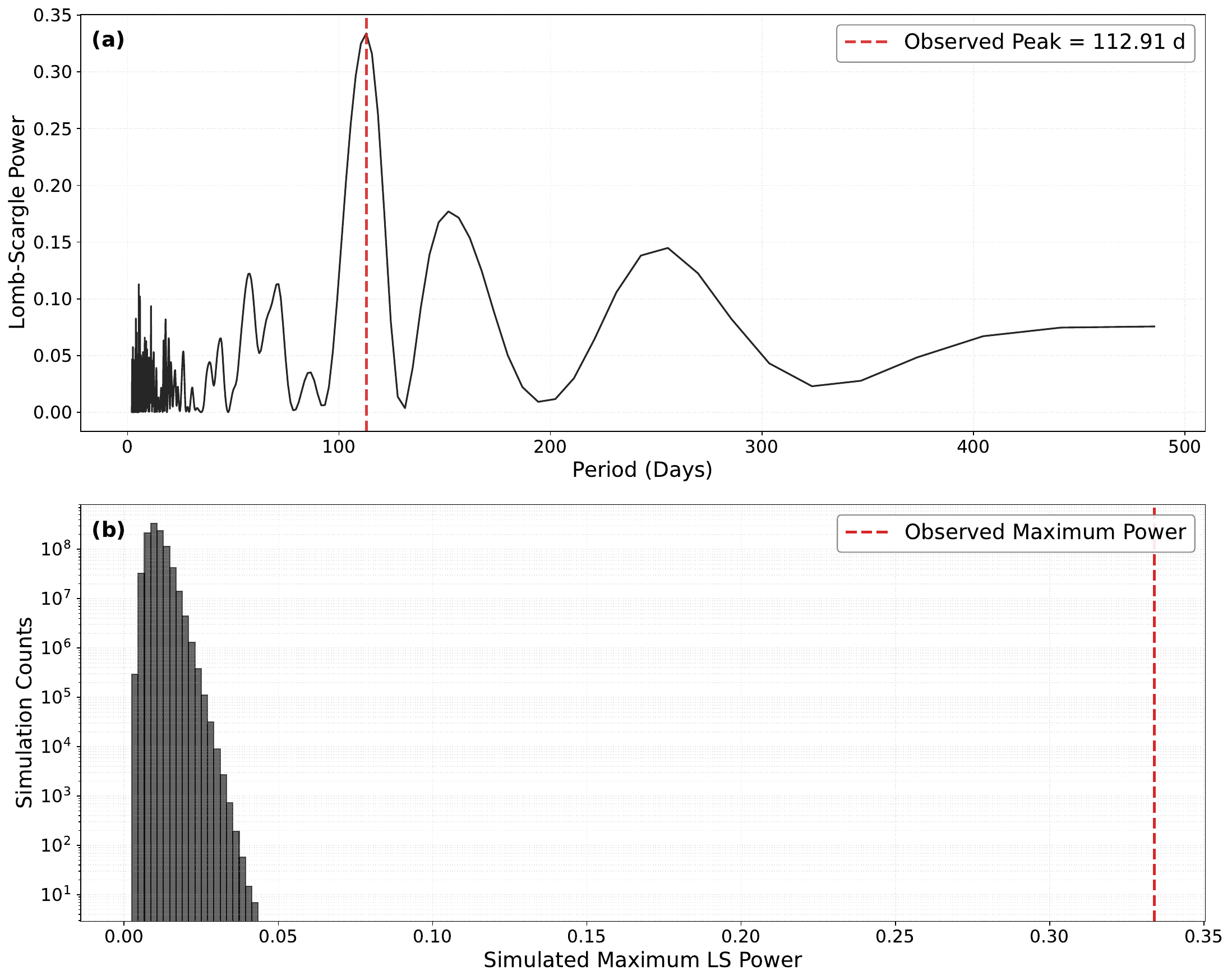}
\caption{Panel (a) shows results of the Lomb-Scargle periodogram of the central frequency for FRB 20240114A. The red dashed line indicates the location of the most significant peak at $112.91$ days. Panel (b) shows the distribution of maximum Lomb-Scargle powers generated from the bootstrap of $10^9$ iterations. The red dashed line marks the maximum power derived from the real data. The FAP and its equivalent Gaussian significance are determined by the fraction of simulated maximum powers that exceed the observed value.} 
\label{Fig1}
\end{figure}

\begin{figure} 
\centering
\includegraphics[width=120 mm]{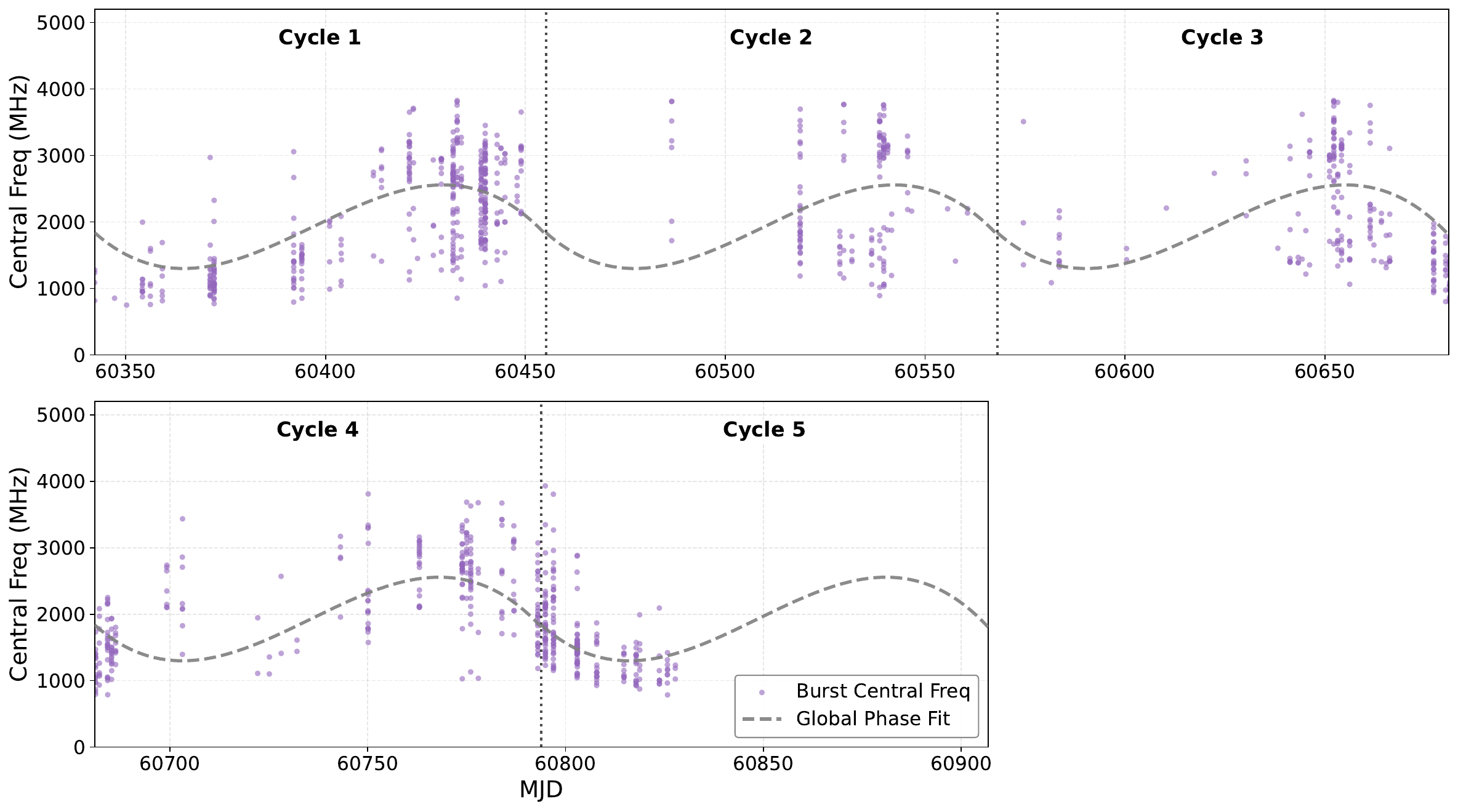}
\caption{Evolution of the observed central frequency for FRB 20240114A across consecutive cycles, based on the identified period of $112.91$ days. Purple circles represent the central frequencies of individual bursts derived by \cite{Uttarkar2026}. The gray dashed line represents an empirical third-order polynomial fit applied to the global phase data, serving as a visual reference to illustrate the consistent evolutionary trend across all cycles. The fitting function is given by $\nu(\phi) = p_0 \phi^3 + p_1 \phi^2 + p_2 \phi + p_3$, where $\phi \in [0, 1)$ is the folded phase.}
\label{Fig2}
\end{figure}

By normalizing the spectral power for the $N$ total bursts, the algorithm quantifies the likelihood of periodicity and determines the most prominent candidate. We applied the Lomb-Scargle tool from the open-source package \texttt{astropy} to search for periodicities in the central frequencies. The search range was set from 2 days up to $T_{\rm span}/3$, because any period with fewer than 3 full cycles lacks statistical significance \citep{Vaughan2016}. As shown in Figure \ref{Fig1}, only one prominent peak was identified, corresponding to a period of 112.91 days with an LS power of ~0.33. Figure \ref{Fig2} illustrates the cycle-by-cycle evolution of the central frequency folded with the 112.91-day period. An empirical third-order polynomial is used to provide a clear visual reference for the phase coherence. Within each period, the central emission frequency is initially low and gradually increases. This frequency periodicity remains phase-coherent across approximately five cycles. 

To evaluate the statistical significance of this period, we employed the bootstrap resampling procedure. Based on $10^8$ simulated runs, the false alarm probability (FAP) is $<10^{-8}$, which is equivalent to a significance of $> 6 \sigma$. By fitting the high power tail of the complementary cumulative distribution function of the simulated maximum LS powers, we derive the extrapolated FAP of $4.61\times10^{-88}$ and equivalent Gaussian significance of $19.86 \sigma$. The calculated FAP derived from the \texttt{false alarm probability} method of the \texttt{LombScargle} class is $10^{-103}$, which corresponds to a significance of $21.58\sigma$. This theoretical FAP was estimated using the Baluev method \citep{Baluev2008}, which is based on extreme value theory. Because this FAP is too small to be fully verified computationally via the bootstrap method, bootstrapping instead yields a conservative lower limit of $> 6\sigma$. We also evaluate the statistical significance of the detected periodicity by employing tail extrapolation based on the extreme value theory \citep{Suveges2014} due to the tail of the maximum power distribution falls off exponentially \citep{Baluev2008}. We modeled the extreme tail (top $1\%$) of our simulated maximum powers using a linear function: $\log_{10}(\text{FAP}) = a Z + b$, where $Z$ is the LS power. The fitting results are shown in panel (a) of Figure \ref{Fig3}. The fitted parameters are $a = -270.41$ and $b = 2.95$. The derived FAP and Gaussian significance are $4.61\times10^{-88}$ and $19.86 \sigma$, respectively. 

\subsection{Phase-folding Method}
To test the robustness of the above result, we also performed the phase-folding analysis to the central frequency data.
For a given trial period, the arrival times were folded into phase bins, and a $\chi_{freq}^2$ statistic was calculated to quantify the deviation of the bin-averaged frequencies from the global mean. The $\chi_{freq}^2$ statistic is defied as
\begin{equation}
\begin{aligned}
\chi_{freq}^2= \sum_{i=1}^{M} \frac{N_i (\bar{\nu}_i - \bar{\nu}_{global})^2}{\sigma_{global}^2}
\end{aligned}
\end{equation}
where $M$ is the total number of valid bins, and $N_i$ and $\bar{\nu}_i$ are the number of data points and the mean central frequency in the $i$-th bin, respectively. $\bar{\nu}_{\mathrm{global}}$ represents the global mean of the central frequency, and $\sigma_{\mathrm{global}}$ is the global standard deviation of the sample. To maintain statistical reliability, we required each valid phase bin to contain a minimum of two data points.

The trial period yielding the maximum $\chi_{freq}^2$ represents the most plausible candidate. As illustrated in Figure \ref{Fig4}, the periodogram reveals two highly significant peaks at 112.28 and 223.40 days. The latter corresponds to a harmonic of the primary 112.28-day period. This finding is consistent with the LS result.

To ensure that the detection is robust and not an artifact of the chosen binning resolution, we systematically tested the impact of varying the number of bins on the results. As shown in Figure \ref{Fig5}, varying the number of phase bins from 5 to 20 does not significantly alter the identified period. The plausible period range spans 104–123 days with a mean value of 113.5 days, which is in excellent agreement with the 112.91-day period identified via the LS method. These results confirm that the identified periodicity is statistically robust.

\begin{figure} 
\centering
\includegraphics[width=120mm]{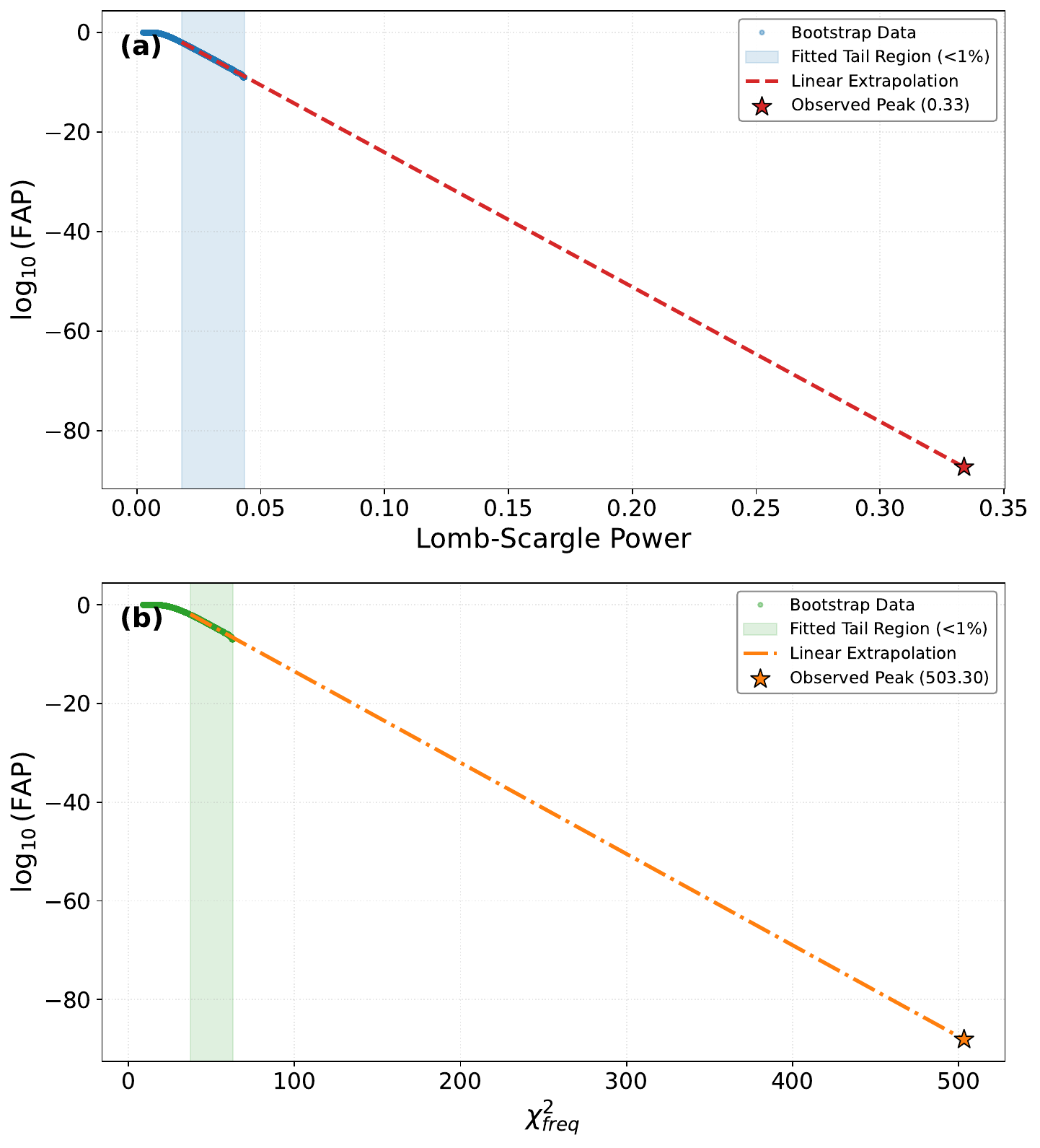}
\caption{Panel (a) is the extrapolation diagnostic plot of Lomb-Scargle periodogram. The blue circles represent the empirical CCDF of the maximum LS powers derived from $10^9$ bootstrap simulations. The red dashed line shows the optimal linear fit to the top 1 $\%$ of the simulated distribution. The red star indicates the actual observed peak power ($Z_{\rm obs} = 0.33$). The extrapolated FAP of $3.17\times10^{-87}$ corresponds to an equivalent Gaussian significance of $19.76 \sigma$. Panel (b) is the extrapolation diagnostic plot of phase-folding periodogram. The green circles represent the empirical CCDF of maximum $\chi_{freq}^2$ derived from $10^9$ bootstrap simulations. The red orange line shows the optimal linear fit to the top 1 $\%$ of the simulated distribution. The orange star indicates the actual observed peak power ($Z_{\rm obs} = 503.30$). The extrapolated FAP of $6.76\times10^{-90}$ corresponds to an equivalent Gaussian significance of $20.07 \sigma$.}
\label{Fig3}
\end{figure}

\begin{figure} 
\centering
\includegraphics[width=120 mm]{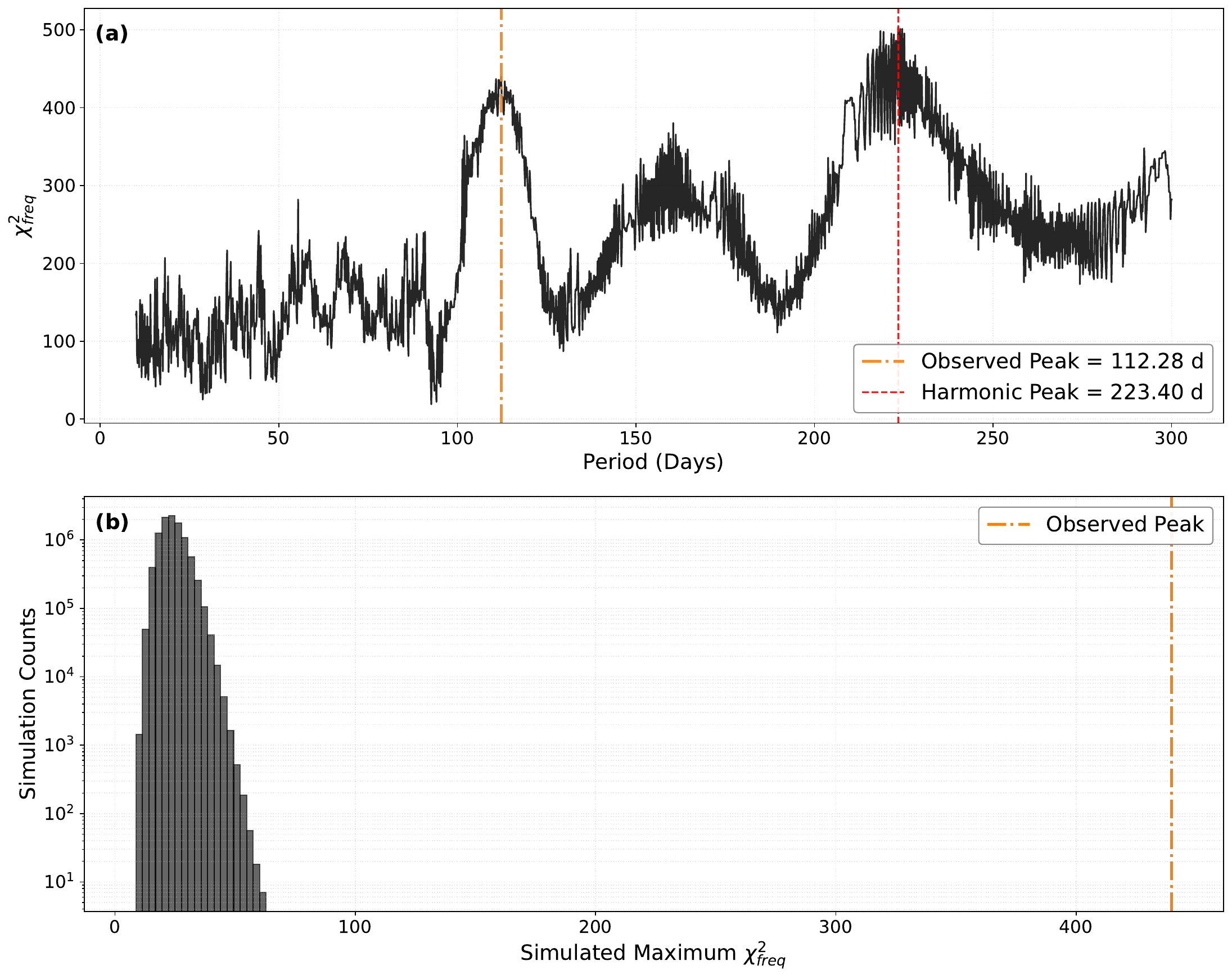}
\caption{Panel (a) represents the results of phase-folding periodogram of the central frequency for FRB 20240114A with 10 phase bins. The orange dot-dashed line indicates the location of the significant peak at $112.28$ days, while the red dotted line indicates the harmonic peak at 223.40 days. Panel (b) shows the distribution of $\chi_{freq}^2$ generated from the bootstrap of $10^9$ iterations. The orange dashed line marks the maximum power observed in the real data. The FAP and its equivalent Gaussian significance are determined by the fraction of simulated maximum powers that exceed the observed value.}
\label{Fig4}
\end{figure}

\begin{figure} 
\centering
\includegraphics[width=120 mm]{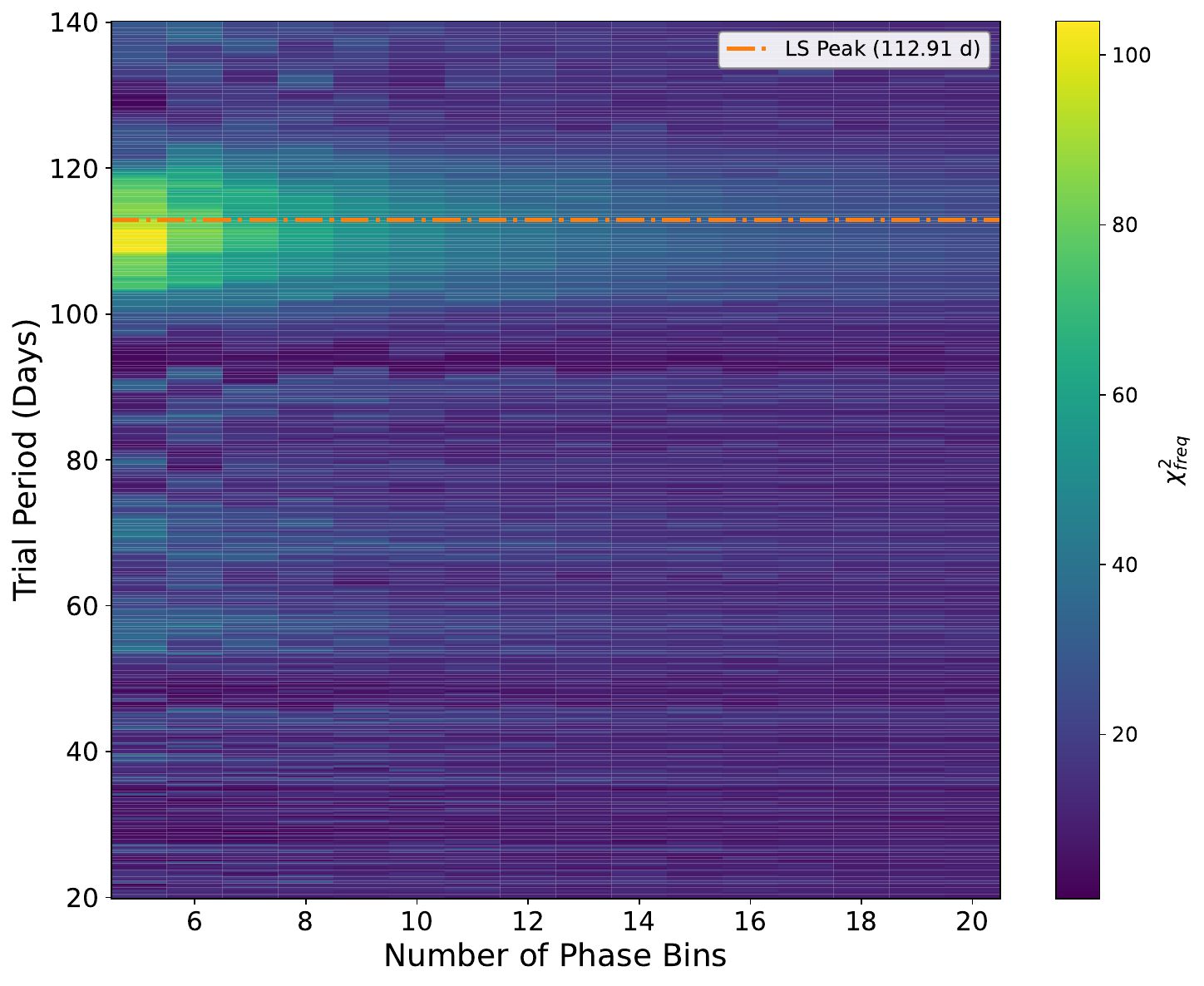}
\caption{Significance of the phase-folding analysis across different phase bin sizes and trial periods. The color scale represents the $\chi_{freq}^2$ statistic, which reflects the detection significance for each parameter combination. The most significant periodic signal strongly peaks at $\sim 112$ days.}
\label{Fig5}
\end{figure}

Furthermore, we conducted a Monte Carlo permutation test comprising $10^9$ iterations using 10 phase bins. The phase-folding pipeline was applied to each shuffled dataset to record the maximum $\chi_{freq}^2$, constructing an empirical null distribution. By comparing the observed maximum $\chi_{freq}^2$ against this simulated distribution, we derived a final FAP of $<10^{-9}$, corresponding to a lower limit on the Gaussian significance of $>6\sigma$. Analogous to the Lomb-Scargle analysis, we also evaluated the statistical significance of the detected periodicity by employing tail extrapolation based on extreme value theory \citep{Suveges2014}. We modeled the top $1\%$ of our simulated maximum scores using a linear function: $\log_{10}(\text{FAP}) = a Z + b$, where $Z$ is the $\chi_{freq}^2$ of the phase-folding periodogram. The fitting results are presented in panel (b) of Figure \ref{Fig3}. The fitted parameters are $a = -0.19$ and $b = 5.11$, yielding an extrapolated FAP of $6.76 \times 10^{-90}$ and an equivalent Gaussian significance of $20.07\sigma$.

\section{Possible interpretations} \label{sec:theory}
In this section, we investigate the physical mechanisms driving the periodic modulation in the central frequency of FRB 20240114A, including the intrinsic radiation mechanism, extrinsic propagation effect and the geometrical modulation. 

\subsection{Intrinsic-Origin Interpretation}
Coherent curvature radiation by bunches in the magnetosphere has been regarded as one of the plausible radiation mechanisms for FRBs \citep{Katz2014,Wang2016,Kumar2017,Yang2018}. The characteristic frequency of curvature radiation is $v_{CR}\sim(\rm{0.72~GHz})\gamma_{e,2}^3\rho_7^{-1}$, where $\gamma_e$ is the electron Lorentz factor and $\rho$ is the curvature radius \citep{Zhang2023Review}. The periodic modulation induced solely by curvature radiation process requires periodic modulation in $\gamma_e$ or $\rho$ which is unlikely due to the lack of the corresponding observational evidence in normal NS. 

Alternatively, coherent inverse Compton scattering (ICS) by relativistic particle bunches has been proposed to account for various observational features of FRBs \citep{Zhang2022_ICS, Qu2024}. In this framework, the frequency of the scattered radiation can be parameterized as $\nu \simeq (1 \text{ GHz}) \gamma_{2.9}^2 \omega_{i,4}$, where $\omega_i$ represents the frequency of the incident fast magnetosonic waves. Attributing the macroscopic 112-day spectral drift to periodic variations in the incident wave frequency is physically disfavored, given that these waves are expected to be excited by stochastic crustal yielding operating on millisecond dynamical timescales \citep{Burnaz2025, Qu2026}. Moreover, because the dependences of the emission frequency on both the incident and scattered angles are intricately coupled, invoking purely angular variations within the ICS model struggles to naturally reproduce the strictly periodic modulation observed in FRB 20240114A.

In the synchrotron maser emission mechanism framework, the characteristic peak frequency of the FRB relies on the shock Lorentz factor $\Gamma$ and the upstream medium density $n_u$, scaling as $\nu \propto \Gamma \sqrt{n_u}$ \citep{Metzger2019, Beloborodov2020}. For an isolated magnetar, both the flare energy that determines $\Gamma$ and the pre-existing wind density $n_u$ are dictated by the stochastic outburst history of the source. Consequently, it is physically unviable to attribute a 112-day periodicity to such random intrinsic activities. 

\subsection{Extrinsic-Origin Interpretation}
In this section we consider the extrinsic-origin for the periodic modulation. We assume that the underlying FRB trigger mechanism does not exhibit any long-timescale periodicity. The characteristic spin timescales of NSs typically range from milliseconds to seconds \citep{Kaspi2017}. Given this stark contrast in timescales, it is challenging to naturally attribute a 112-day long-term periodicity solely to intrinsic magnetospheric or crustal activities. Therefore, A more plausible scenario is that the observed periodic evolution of the central frequency are primarily modulated by extrinsic frequency-dependent propagation effects or geometric modulation. 

FRB 20240114A exhibits a variable RM, which could be the imprint of a companion massive star, similar to that is observed in FRB 20190520B, FRB 20201124A, and FRB 20220529A \citep{Anna-Thomas2023, Xu2022, LiYe2026}. 

As proposed by \cite{Uttarkar2026}, the distinct burst activity and spectral evolution of FRB 20240114A may originate from plasma lensing in a turbulent circumburst medium. While plasma lenses can naturally form in diverse environments—such as supernova remnants (SNRs), pulsar wind nebulae (PWNe), or circumstellar structures \citep{Graham2011, cordes2017, Main2018}. The presence of a strict periodic frequency modulation strongly favors a binary companion origin, as other environments cannot easily sustain such long-term periodicity. Indeed, frequency-dependent lensing effect induced by ionized companion winds has been observed in several Galactic binary systems \citep{Backer2000, Main2018, Lin2021, Bilous2019}. In this context, we consider a binary progenitor consisting of an NS orbiting a massive companion (e.g., an OB-type star). The characteristic length scale of density clumps within the companion's stellar wind/disk can be estimated as \citep{Zhao2023}
\begin{equation}
\begin{aligned}
l_{\mathrm{c}} \sim v_{\mathrm{orb}} \Delta t_{\mathrm{f}} &= 7.2 \times 10^{10} \mathrm{~cm}\left(\frac{M_{\mathrm{tot}}}{30 \ M_{\odot}}\right)^{1 / 2} \times\left(\frac{a}{10^{13} \mathrm{~cm}}\right)^{-1 / 2},
\end{aligned}
\end{equation}
where $v_{\mathrm{orb}}$ represents the NS orbital velocity, $\Delta t_{\mathrm{f}}$ is the characteristic fluctuation timescale, $M_{\mathrm{tot}}$ is the total binary mass, and $a$ is the orbital semi-major axis. Typical bulk velocities for the stellar wind ($v_w$) and the Keplerian circumstellar disk ($v_d$) are on the order of $10^8 \mathrm{~cm \ s^{-1}}$ \citep{Snow1981, Krticka2014} and $10^7 \mathrm{~cm \ s^{-1}}$ \citep{Meilland2007}, respectively. Assuming the plasma clumps are embedded within these mediums, the duration for the LoS to traverse a single clump is roughly $t_s \sim l_c / v_w \simeq 10^3~\rm{}s$ or the wind scenario, and $t_s \sim l_c / v_d\simeq 10^4~\rm{}s$ for the disk. Because these crossing times are orders of magnitude shorter than the $\sim 111$-day modulation period, the long-term frequency periodicity cannot be attributed to the transit of an individual clump. Furthermore, during each orbital transit, the LoS may not intersect the exact same plasma lens at precisely the same orbital phase. 

Free-free absorption could induce frequency-dependent modulation in the central frequency of the radio emission. Assuming the periodicity is driven by orbital motion, we consider a binary system scenario where the companion's stellar wind or disk acts as the absorbing medium. The free-free absorption coefficient of a plasma is given by \cite{Rybicki1979}
\begin{equation}\label{equ:ffa}
\begin{aligned}
\alpha_{\nu}^{\text{ff}} = 0.018 T^{-3/2} z_{\text{i}}^{2} n_{\text{e}} n_{\text{i}} \nu^{-2} \bar{g}_{\text{ff}} 
\end{aligned}
\end{equation}
where $T$ is the temperature of the circumstellar medium, $z_i\sim1$ is the atomic number of the ion, $\bar{g}_{\text{ff}}\sim1$ is the Gaunt factor, $n_e$ and $n_i$ are the number of number densities of electrons and ions, respectively. 
The free–free optical depth of the stellar outflow is
\begin{equation}
\begin{aligned}
\tau_{ff}(\nu) = \int_{\text{LoS}} \alpha^{\text{ff}}(\nu) \, \mathrm{d}l.
\end{aligned}
\end{equation}
We assume that the emission becomes unobservable at orbital phases where the free-free optical depth exceeds unity ($\tau > 1$). 
However, no emission above 2 GHz of FRB 20240114A is observed during several phases. According to Equation (\ref{equ:ffa}), free-free absorption preferentially attenuates lower frequencies. Therefore, it cannot account for the absence of high-frequency emission when low-frequency emission remains visible. In this case, the free-free absorption scenario is highly unlikely.

Pure absorption alone is unlikely to account for the full phenomenology. In radio pulsars, however, free-free absorption and/or cyclotron-resonant absorption can reshape an underlying spectrum into a gigahertz-peaked spectrum (GPS) \citep{Kijak2007}. In this picture, the absorber reprocesses a pre-existing spectrum into an apparent peak \citep{Rajwade2016}. If the NS resides in a binary system, both the density structure and the magnetic geometry along the line of sight can vary with orbital phase, driving a corresponding evolution in the observed spectrum \citep{Kijak2011}. Therefore, if the high-frequency bursts are intrinsically weaker (i.e., a steep negative-index power law) in the frequency range from $\sim$1 GHz to $\sim$4 GHz, any enhanced broadband attenuation can push the faint high-frequency emission below the telescope's detection threshold. Consequently, during orbital phases of strong environmental absorption, this interplay between the intrinsic spectrum and the sensitivity limit will manifest as an apparent spectral cutoff at high frequencies, while the intrinsically brighter low-frequency emission remains observable.

For FRB 20240114A, there was essentially no detectable emission above 2 GHz at early times. The analogy with GPS pulsars is still useful if the absorber is reshaping not a broadband continuum, but a burst population with a broad distribution of energies and central frequencies. Pulsars are broadband emitters, whereas FRBs are narrowband burst by burst, but the ensemble can still develop a peak-like observed spectrum after propagation through a dense magnetized environment. A generic propagation model can therefore be written as
\begin{equation}
S_{\nu, \text {obs}}(\phi)=S_{\nu, \text {int}}(\phi) \exp \left[-\tau_{\text {ff }}(\nu, \phi)-\tau_{\text {cyc }}(\nu, \phi)\right],
\end{equation}
where $S_{\nu, \text{int}}$ is the burst-to-burst intrinsic flux spectrum, $S_{\nu, \text {obs}}$ is the measured flux, $\phi$ is the orbital phase, $\tau_{\text {ff }}$ and $\tau_{\text {cyc }}$ are the optical depth of the free-free absorption and cyclotron-resonant absorption, respectively. 

This model makes specific predictions for the observed spectrum and polarization. Near cyclotron resonance
\begin{equation}
\nu_{\mathrm{res}}=\nu_B /\left[\gamma\left(1-\beta \cos \theta_B\right)\right],
\end{equation}
the two natural modes can suffer different optical depths, so stronger absorption can lower the total flux, reduce the linear polarization fraction, and generate or enhance circular polarization. 

\begin{figure} 
\centering
\includegraphics[width= 120 mm]{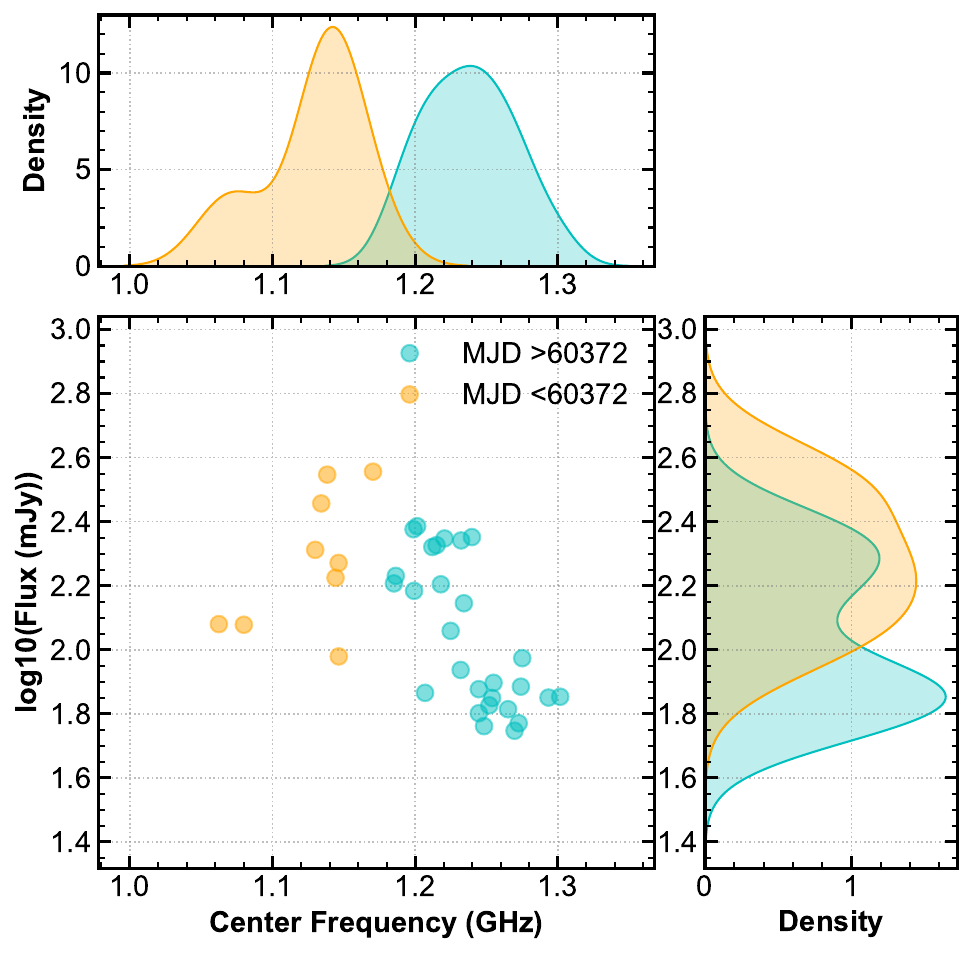}
\caption{Central frequency versus flux density for FAST bursts, together with the corresponding kernel-density distributions. The data is taken from \cite{ZhangJS2025} and the time span is consistent with cycle 1 in Figure \ref{Fig2}. Bursts detected before MJD~60372 are concentrated at lower central frequencies and generally higher flux densities, whereas bursts detected after MJD~60372 shift to higher central frequencies and become systematically fainter.
}
\label{Fig6}
\end{figure}

\begin{figure} 
\centering
\includegraphics[width= 120 mm]{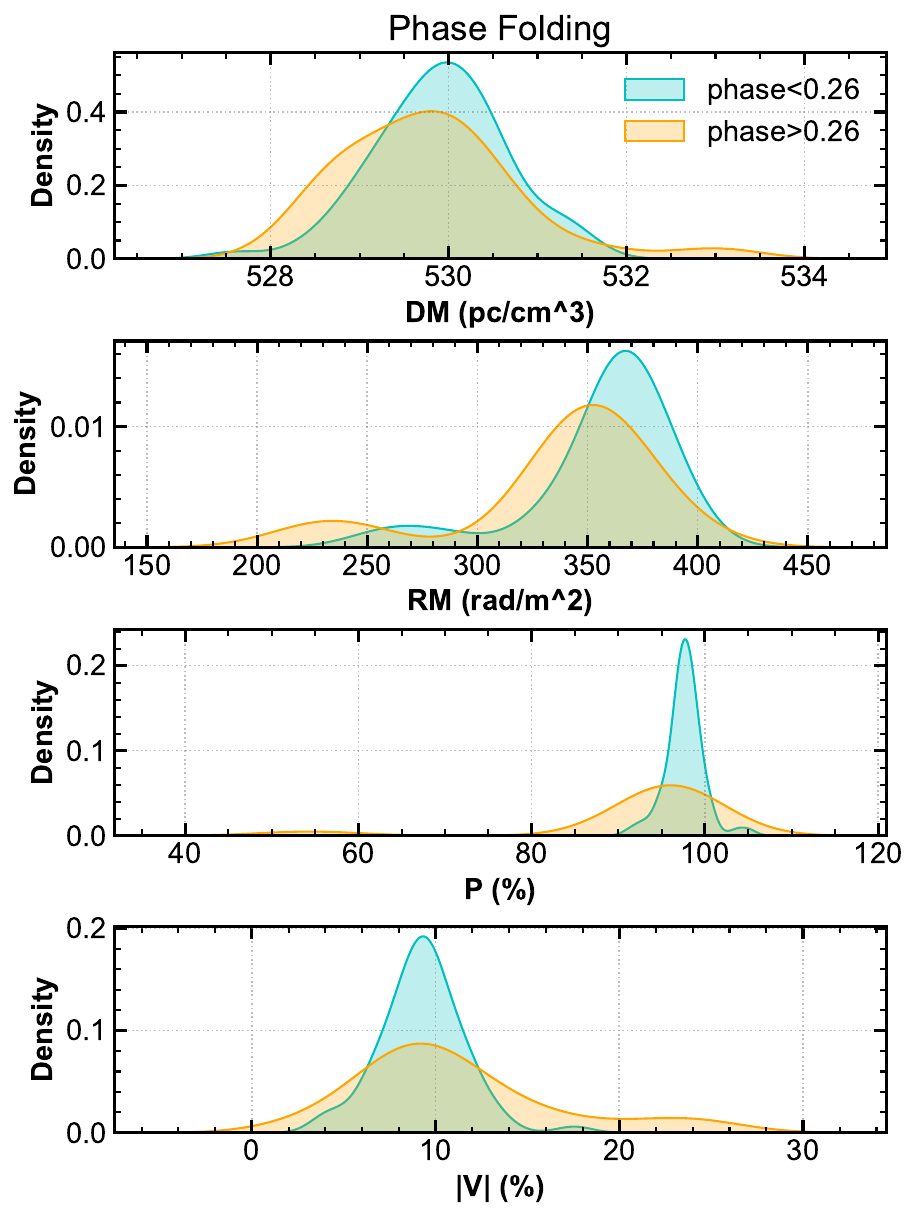}
\caption{The kernel-density distributions of DM, RM, polarization fraction $P$, and circular polarization degree $|V|$. The data is taken from \cite{WangTC2026}. The distributions are phase-folding with a period of 112.91 days, and are separated at phase $0.26$ (MJD~60372). The two phase intervals show only modest differences in DM, but markedly different RM and polarization properties, consistent with phase-dependent propagation through a dense and magnetized environment.}
\label{Fig7}
\end{figure}

To test these expectations, we analyzed contemporaneous FAST observations of FRB 20240114A \citep{ZhangJS2025, ZhangLX2025, WangTC2026, ZhouDK2026}. Central frequency versus flux density for FAST bursts, together with the corresponding kernel-density distributions are shown in Figure \ref{Fig6}. The data is taken from \cite{ZhangJS2025} and the time span is consistent with cycle 1 in Figure \ref{Fig2}. Bursts detected before MJD~60372 are concentrated at lower central frequencies and generally higher flux densities, whereas bursts detected after MJD~60372 shift to higher central frequencies and become systematically fainter. The kernel-density distributions of DM, RM, polarization fraction $P$, and circular polarization degree $|V|$ are shown in Figure \ref{Fig7}. The data is taken from \citep{WangTC2026}. The distributions are phase-folding with a period of 112.91 days, and are separated at phase $0.26$ (MJD~60372). The two phase intervals show only modest differences in DM, but markedly different RM and polarization properties, consistent with phase-dependent propagation through a dense and magnetized environment.

Several theoretical models have been proposed to explain the 16.35-day periodicity of FRB 20180916B \citep{CHIME2020180916}, among which the free or forced precession of an NS, or a slowly rotating NS, are widely discussed \citep{Levin2020, Tong2020, Yang2020, Zanazzi2020, Beniamini2020}. These models predict that the radiation beam periodically sweeps across and eventually completely off the LoS, creating discrete active windows. Consequently, this geometric sweeping should manifest as a high-significance periodicity in burst arrival times. However, as detailed in Section \ref{sec:dis}, our analysis reveals a profound absence of such arrival-time periodicity in FRB 20240114A.

This apparent contradiction can be elegantly resolved if the LoS remains continuously within the radiation beam throughout the entire precession cycle. While forced precession models typically produce symmetric active windows that struggle to explain chromatic phase drifts, a freely precessing NS naturally accommodates an asymmetric emission geometry \citep{LiDZ2021}. Under this continuous viewing scenario of free precession, the periodic cut-off in burst activity is absent, yet the macroscopic spectral evolution is driven by the continuously varying viewing angle.

Within the framework of coherent curvature radiation, the characteristic emission frequency $\nu$ depends on the Lorentz factor $\gamma$ of the particles and the curvature radius $\rho$ of the emission region
\begin{equation}
\begin{aligned}
\nu = \frac{3}{4\pi}\gamma^3\left(\frac{c}{\rho}\right). 
\end{aligned}
\end{equation}
Assuming a stationary magnetosphere with a dipolar field ($\frac{r}{R_0} = C \sin^2 \delta_{\text{em}}$, where $\delta_{\text{em}}$ is the magnetic polar angle of the field line and $r$ is the radial distance), $\rho$ is given by \cite{Yang2018}
\begin{equation}
\begin{aligned}
\rho = \frac{r(1 + 3 \cos^2 \delta_{\text{em}})^{\frac{3}{2}}}{3 \sin \delta_{\text{em}}(1 + \cos^2 \delta_{\text{em}})} \equiv r F(\delta_{\text{em}}).
\end{aligned}
\end{equation}
Following \cite{LiDZ2021}, we adopt an altitude-dependent Lorentz factor $\gamma(r) \propto r^{-2/3}$, which yields $\nu \propto r^{-3}$ \citep{LiDZ2021}. Under the small-angle approximation ($\sin\delta_{\text{em}}\simeq \delta_{\text{em}}$), we have $r \propto \delta_{\text{em}}^2$. Furthermore, the observed magnetic polar angle $\delta_{\text{ob}}$ is related to the emission angle $\delta_{\text{em}}$ by \cite{LiDZ2021}
\begin{equation}
\begin{aligned}
\cos \delta_{\text{ob}} = \frac{1 + 3 \cos 2\delta_{\text{em}}}{\sqrt{10 + 6 \cos 2\delta_{\text{em}}}}.
\end{aligned}
\end{equation}

In the small-angle regime, $\delta_{\text{ob}} \propto \delta_{\text{em}}$, leading to $\delta_{\text{ob}} \propto r^{1/2}$. Combining this with the frequency-altitude relation, we obtain the scaling law between the observed polar angle and the central frequency: $\delta_{\text{ob}} \propto \nu^{-1/6}$.

Observationally, the central frequency of FRB 20240114A varies from $\sim 3$~GHz down to $\sim 1$~GHz within one cycle. According to the scaling law, this macroscopic spectral drift requires a fractional change in the viewing angle of $\delta_{\text{ob}}(1\text{ GHz}) / \delta_{\text{ob}}(3\text{ GHz}) = 3^{1/6} \simeq 1.2$. If we assume a reference impact angle of $\delta_{\text{ob,ref}} = 10^\circ$, the required wobble angle induced by free precession is only $\sim 0.9^\circ$. As illustrated in \citep{LiDZ2021}, the emission region for a specific frequency forms a cone centered at $\hat{m}_f$ with an intrinsic opening angle of $\xi_f$. Because this tiny geometric wobble ($\sim 0.9^\circ$) is significantly smaller than typical beam opening angles ($\xi_f \gtrsim 5^\circ$), the LoS naturally remains trapped within the active radiation beam. This quantitative consistency supports that a freely precessing magnetar can reproduce the frequency periodicity without producing the arrival-time periodicity. Although free precession implies a varying viewing geometry, the lack of a periodic PA modulation in FRB 20240114A is consistent with our model. The tiny wobble angle (e.g. $\sim 0.9^\circ$) yields negligible PA variation within the rotating vector model \citep{Radhakrishnan1969}. Furthermore, any minor underlying trend would be obscured by intrinsic pulse-to-pulse PA jitter, and local RM fluctuations \citep{Cordes1978, Luo2020, Feng2022}.

Alternative to magnetar engines, the periodic modulation of FRBs can also be governed by Lense-Thirring (LT) precession of a super-Eddington accretion disk in an X-ray binary system \citep{Katz2020b, Sridhar2021}. In this scenario, FRB is produced by relativistic shocks propagating into a pre-existing structured jet \citep{Sridhar2021}. As the jet precesses, the LoS intersects different angular regions of the structured jet, which naturally yields a periodic macroscopic drift in the central frequency. However, super-Eddington accretion inherently produces a geometrically thick disk with a narrowly collimated polar funnel \citep{Sridhar2021}. Similar to free-free precession model, the LoS must remain continuously confined within this extremely narrow funnel throughout the LT precession cycle to avoid producing the periodic activity. While possible, satisfying this stringent continuous viewing constraint, while simultaneously traversing a sufficient gradient in the structured jet to drive a large $1-3$~GHz spectral drift requires significant fine-tuning geometric parameter. In contrast, the relatively wider intrinsic emission beams characteristic of the magnetar coherent curvature radiation \citep{LiDZ2021} accommodate this continuous-viewing geometry much more naturally. 

\section{Discussion} \label{sec:dis}
\subsection{Co-evolution in FAST Data}
\begin{figure} 
\centering
\includegraphics[width=120 mm]{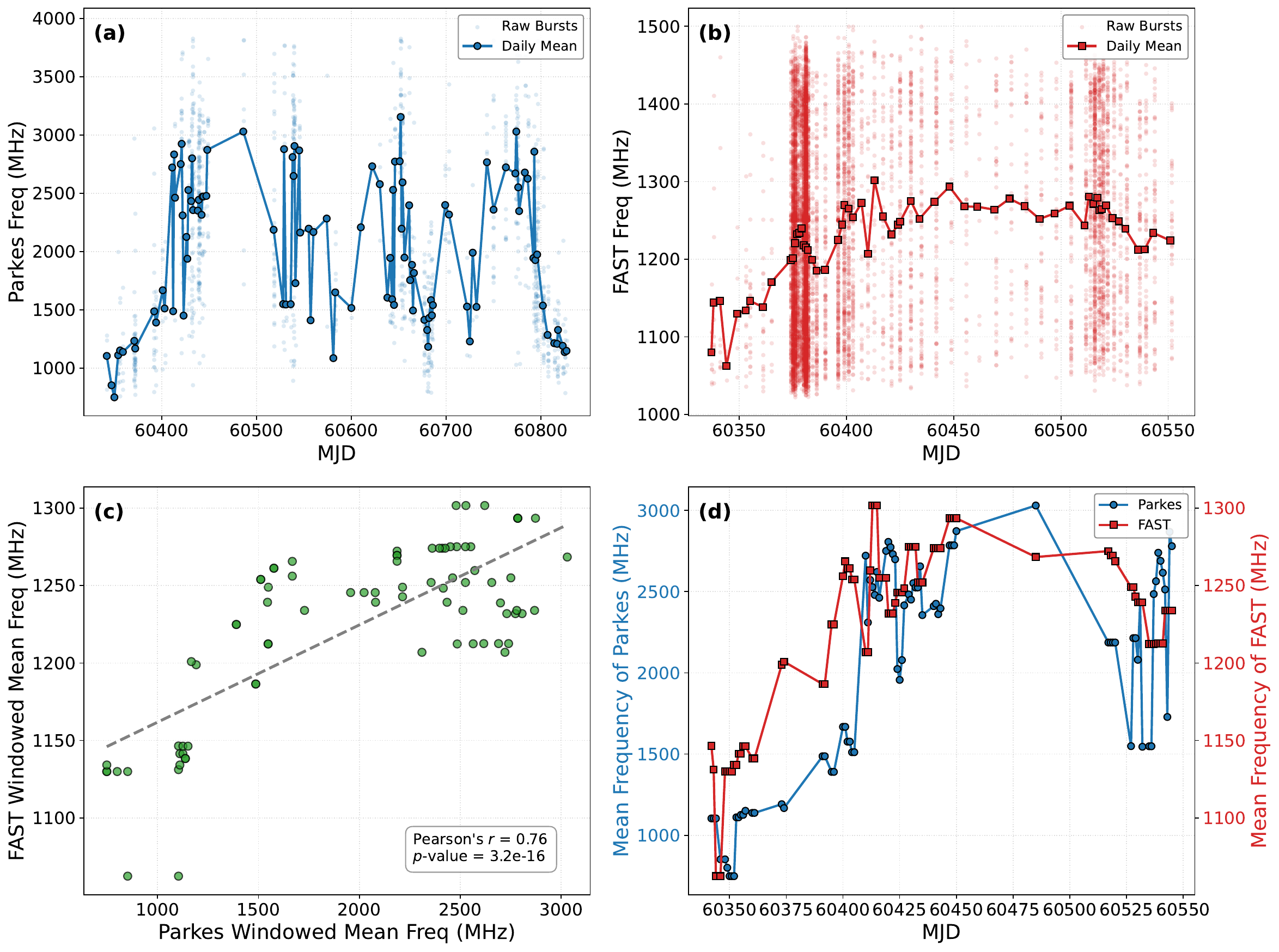}
\caption{Panel (a) illustrates the central frequency evolution observed by the Parkes UWL. Light blue dots represent individual raw bursts, while solid blue circles trace the daily mean frequencies. Panel (b) shows the same as panel (a), but for FAST observations, utilizing light red dots for raw bursts and solid red squares for daily means. Panel (c) displays the correlation between the windowed mean frequencies (using a $\pm2$-day sliding window) of Parkes and FAST during overlapping epochs. The dashed gray line represents the best linear fit, demonstrating a significant positive correlation (Pearson $r = 0.76$, $p < 10^{-15}$). Panel (d) presents the evolution of the daily mean central frequencies. The dual Y-axis visualization highlights the synchronized long-term frequency variations detected independently by both telescopes.}
\label{Fig8}
\end{figure}

Here, we show that burst samples from both Parkes and FAST exhibit a consistent evolutionary trend of the central frequency.
FAST monitored FRB 20240114A from MJD 60337 to 60825, fully encompassing our Parkes observational epoch \citep{WangTC2026}. Notably, central frequencies are only available for bursts prior to MJD 60552 \citep{ZhangJS2025}. To account for bursts truncated by the FAST bandpass edges, a frequency-extended Gaussian fit was applied to correct the bandwidth, while a boxcar bandwidth was used for faint or morphologically complex bursts. Panel (a) and (b) of Figure \ref{Fig8} present the central frequency of each burst as a function of MJD for the Parkes and FAST datasets, respectively. A visual inspection reveals a potential co-evolutionary trend between the two samples. To quantify this correlation, we performed a statistical test as shown in panel (c) of Figure \ref{Fig8}, yielding a Pearson correlation coefficient of $0.76$, which strongly indicates a strongly correlated evolution. Panel (d) of Figure \ref{Fig8} further illustrates these evolutionary trends after scaling the y-axis for both data samples. Nevertheless, the central frequency estimates from FAST are generally less reliable than those from Parkes due to its limited instrumental bandpass.

\subsection{Searching for Periodic Activity}
\begin{figure} 
\centering
\includegraphics[width=120 mm]{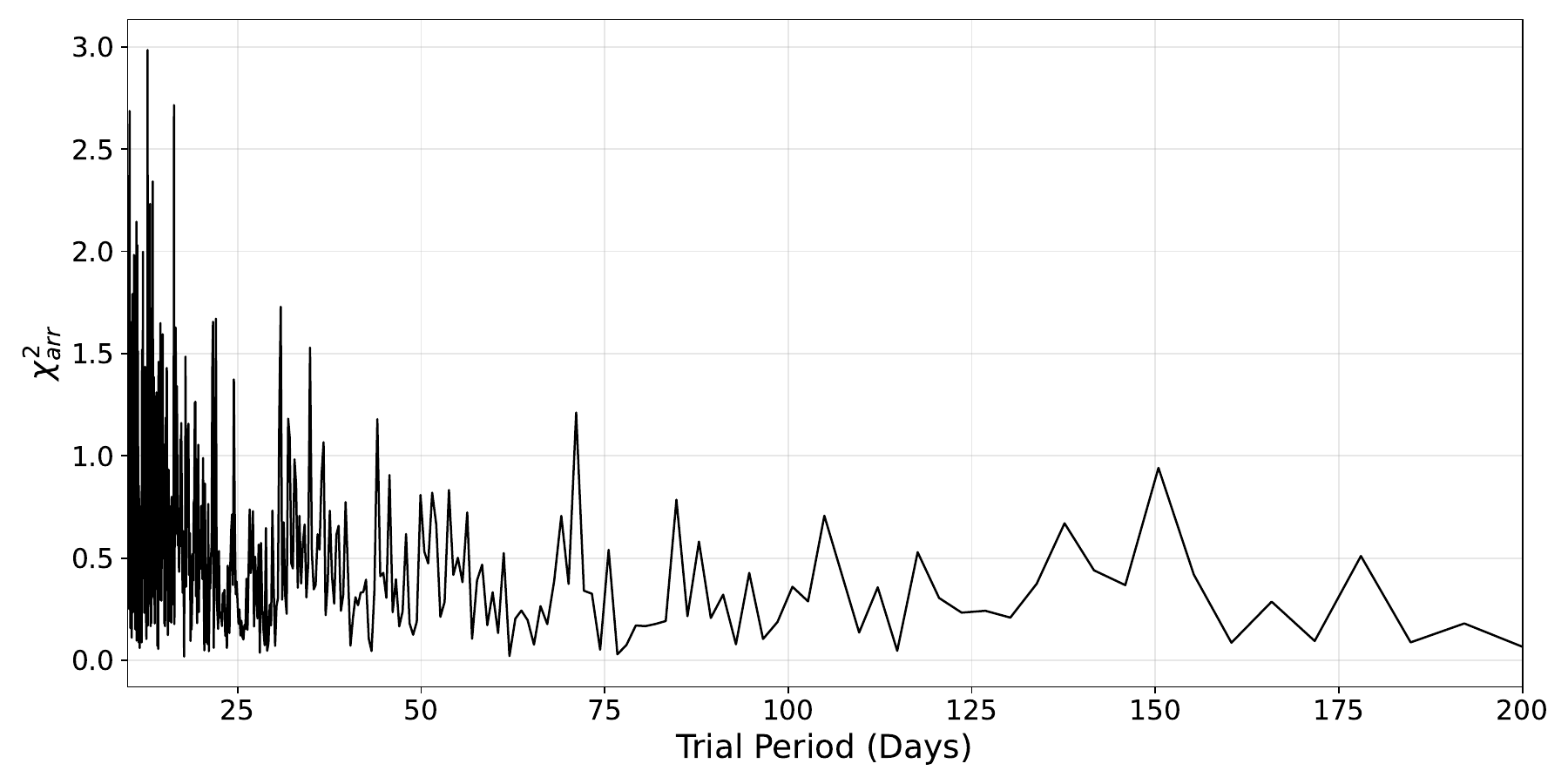}
\caption{The $\chi_{arr}^2$ periodogram with respect to a uniform distribution of burst arrival times for different folding periods for bursts with S/N $>$ 20. Only bursts separated by a sidereal day are considered independent in this approach. No significant frequency peak is identified.}
\label{Fig9}
\end{figure}

We test whether the central frequency modulation caused by the burst activity. To search for periodicity in the burst arrival times of this sample, we fold the arrival times with different periods and group the folded bursts into $n$ phase bins. Following \citep{CHIME2020180916}, an exposure-weighted Pearson $\chi^2$ test for deviation from uniformity was employed
\begin{equation}\label{equ:3}
\chi_{arr}^2 = \sum_{i=1}^{n} \frac{(N_i - E_i)^2}{E_i},
\end{equation}
where $n$ is the total number of phase bins, $N_i$ is the observed number of active days in bin $i$, $E_i = p T_i$ is the expected number of active days from a uniform distribution, $T_i$ is the integrated exposure time for bin $i$, and $p = \sum N_i / \sum T_i$ is the average active-day rate per unit exposure time. As shown in Equation (\ref{equ:3}), the expectation in each phase bin is strictly weighted by the accumulated exposure time to account for the uneven observational scheduling. The exposure times for individual observations are detailed in Extended Data Table 1 of \cite{Uttarkar2026}. The active days and their corresponding telescope exposure times were folded into $n$ phase bins over a grid of trial periods. The trial frequencies were uniformly sampled with a resolution of $0.1/T_{\rm span}$, where $T_{\rm span} \sim 485$ days is the total observational baseline. Following \citep{CHIME2020180916}, we assigned a total weight of unity to bursts that arrived on the same sidereal day to avoid contamination from short-timescale correlated processes. The $\chi_{arr}^2$ as a function of the trial period, calculated using five phase bins, is shown in Figure \ref{Fig9}. No statistically significant peak is observed. This is consistent with the null result of the periodicity search reported by \cite{Uttarkar2026}, which utilized bursts with a S/N above 8. Furthermore, varying the number of phase bins does not alter the main conclusion of this analysis.

\section{Summary} \label{sec:summary}
In this work, we applied Lomb-Scargle (LS) periodogram and phase-folding techniques to analyze the burst central frequencies, identifying a prominent periodicity at 112.91 days. To rigorously evaluate the statistical significance of this detection, we performed $10^8$ bootstrap resampling simulations. The resulting FAP is $<10^{-8}$, corresponding to a significance of $>6\sigma$. By extrapolating the high-power tail of the complementary cumulative distribution function, we derive equivalent Gaussian significances of $19.86\sigma$ and $20.07\sigma$ for the LS and phase-folding methods, respectively. We explored potential physical mechanisms driving this periodic frequency modulation. Both phase-dependent absorption within a massive binary system and the free precession of a magnetar emerge as viable interpretations. Alternatively, the observed modulation may arise from a complex interplay of intrinsic trigger mechanisms, local radiation physics, and geometric propagation effects. Future continuous monitoring of active FRBs with ultra-wideband receivers will be crucial in disentangling these scenarios and unveiling the precise radiation mechanisms.

\clearpage

\section*{Acknowledgments}
Rui-Nan Li acknowledges the helpful discussion with Yue Wu, Chen-Ran Hu, Tian-Cong Wang and Jun-Shuo Zhang. 
This work was supported by the National Natural Science Foundation of China (grant Nos. 12494575, 12393812 and 12273009). Qin Wu is supported by the National Natural Science Foundation of China (grant Nos. 12447115 and 12503050) and the China Postdoctoral Science Foundation under Grant Number GZB20240308, 2025T180875 and 2025M773199.

\bibliographystyle{aasjournal}
\bibliography{ref}

\clearpage

\end{document}